\documentclass[]{aa}
\usepackage{graphicx}
\usepackage{txfonts}
\usepackage{natbib}

\bibpunct[]{(}{)}{;}{a}{}{,}
\def\ltsima{$\; \buildrel < \over \sim \;$}
\def\simlt{\lower.5ex\hbox{\ltsima}}
\def\gtsima{$\; \buildrel > \over \sim \;$}
\def\simgt{\lower.5ex\hbox{\gtsima}}

\begin{document}
   \title{Disappearance of N$_2$H$^+$ from the Gas Phase \\
          in the Class 0 Protostar IRAM~04191\thanks{Based on observations 
          carried out with the IRAM Plateau de Bure Interferometer. IRAM is 
          supported by INSU/CNRS (France), MPG (Germany) and IGN (Spain).}}

   \author{A. Belloche
          \inst{1,2}
          \and
          P. Andr\'e\inst{3}
          }

   \offprints{A. Belloche}

   \institute{Max-Planck-Institut f\"ur Radioastronomie, Auf dem H\"ugel 69, 
              D-53121 Bonn, Germany\\
              \email{belloche@mpifr-bonn.mpg.de}
         \and
              LERMA/LRA, Ecole Normale Sup\'erieure, 24 rue Lhomond, 
              F-75231 Paris Cedex 05, France
         \and
              Service d'Astrophysique, CEA/DSM/DAPNIA, C.E. Saclay,
              F-91191, Gif-sur-Yvette Cedex, France\\
              \email{pandre@cea.fr}
              }

   \date{Received 23 February 2004; accepted 10 April 2004}

   \abstract{We present a high-resolution millimeter study of the very young 
   Class~0 protostar IRAM~04191+1522 in the Taurus molecular cloud. 
   N$_2$H$^+$(1-0) observations
   with the IRAM Plateau de Bure Interferometer and 30m telescope 
   demonstrate that the molecular ion N$_2$H$^+$ disappears from the gas phase
   in the inner part of the protostellar envelope ($r < 1600$ AU, 
   $n_{\mbox{\tiny {H$_2$}}} > 5 \times 10^5$ cm$^{-3}$). This result 
   departs from the predictions of current chemical models. It suggests 
   either that N$_2$ is more depleted than the models predict, owing 
   to a higher binding energy on polar ice or an enhanced grain chemistry 
   transforming N$_2$ to less volatile species, or that strong deuterium 
   fractionation enhances N$_2$D$^+$ to the detriment of N$_2$H$^+$.
   
   }

   \maketitle
%

\section{Introduction}
\label{sec_intro}

Understanding the onset of gravitational collapse in dense cloud cores 
requires detailed (sub)millimeter studies of the structure of 
prestellar condensations and young protostars \citep*[e.g.][]{Andre00}. 
However, recent observations have shown that molecules such as CO and CS 
deplete onto grain surfaces in the inner parts of dense cores 
\citep[e.g.][]{Bacmann02,Tafalla02}. Since this depletion phenomenon 
is thought to affect many other species at high densities, 
studying the kinematics of pre/protostellar cores with 
future high-resolution instruments such as ALMA may be 
difficult and requires the identification of the best tracers of dense gas. 
N$_2$H$^+$ has been put forward as such a good tracer since it is 
much less sensitive to depletion effects than other species observed with
single-dish telescopes \citep[e.g.][]{Tafalla02}, even though \citet{Bergin02} 
reported a slight decrease of its abundance toward the prestellar core B68.

Prime targets for observational studies of protostellar collapse  
are very young Class~0 objects such as 
IRAM~04191+1522 -- hereafter IRAM~04191 for short -- in Taurus
\citep[see][ -- hereafter \citeauthor*{Andre99}]{Andre99}. 
This protostar features a prominent ($\sim 0.5-1.5\, M_\odot$) envelope 
and a powerful bipolar outflow, but lacks a sizeable accretion disk.
\citet[][ -- hereafter \citeauthor*{Belloche02}]{Belloche02}
showed that the envelope is undergoing both
extended infall motions and fast, differential rotation. They proposed that
the rapidly rotating inner envelope ($r <$ 3500 AU) 
corresponds to a magnetically supercritical core decoupling from an
environment still supported by magnetic fields and strongly affected by
magnetic braking.

Here, we report new observations of IRAM~04191 carried out with the 
Plateau de Bure interferometer (PdBI) in the N$_2$H$^+$(1-0) line
in an effort to probe the inner structure of the envelope. 
We discuss the results of these 
high-resolution observations which show that N$_2$H$^+$ disappears
from the gas phase in the inner part of the envelope.


\section{Observations and data reduction}
\label{sec_obs}

We observed IRAM~04191 with the PdBI for 34~h between December 2001 and 
September 2002 with 4 to 6 antennas in the C and D configurations. 
In the 3mm band, each receiver was tuned to the N$_2$H$^+$(101-012) line at 
93.176258 GHz in single side-band mode. The system temperatures were 
typically 110-200 K. The 39 kHz channel spacing resulted in a velocity 
resolution of 0.20 km~s$^{-1}$ \citep[see][]{Wiesemeyer01}. The 
(naturally-weighted) synthesized half-power beam width was $5.8\arcsec 
\times 4.0\arcsec$ (810 AU $\times$ 560 AU) and the (FWHM) primary beam 
$\sim 54\arcsec$.
Several nearby phase calibrators were observed to determine the time-dependent 
complex antenna gains. The correlator bandpass was calibrated on the 
sources 3C84, 0415+379, 3C273, 3C454.3 or NRAO150, while the absolute flux 
density scale was derived from MWC349 or 3C84.  The absolute calibration 
uncertainty is estimated to be $\sim 15 \%$. The data were calibrated and 
imaged using the GILDAS software (Jan. 2002 
release). The deconvolution was performed with the CLEAN method, down to 0.6 
times the rms noise level.

The present study also uses single-dish N$_2$H$^+$(1-0) and 1.3mm continuum 
on-the-fly maps taken with the IRAM 30m telescope  
\citep[see \citeauthor*{Belloche02} and][ -- hereafter 
\citeauthor*{Motte01}]{Motte01}, as well as 1.3mm continuum 
interferometric data obtained with PdBI in 1999 (see \citeauthor*{Belloche02}).
The combination of the PdBI and single-dish data was performed in the 
\textit{uv}-plane \citep[see][]{Guilloteau01}. We used a single-dish 
calibration factor of 5.9 Jy~beam$^{-1}$~K$^{-1}$ for N$_2$H$^+$(1-0)
\citep*[][]{Greve98}. For the continuum data, we applied a 
calibration factor of 1.35
to account for the difference in observed 
frequency (227 GHz at PdBI and $\sim 250$~GHz for bolometer observations 
at the 30m).

\section{Analysis: A drop in N$_2$H$^+$ abundance} 
\label{sec_n2h+_pdbi30m}

\subsection{A central hole of N\/$_2$H\/$^+$ emission}
\label{sub_results}

\begin{figure}[!t]
 \centerline{\resizebox{0.9\hsize}{!}{\includegraphics[angle=270]{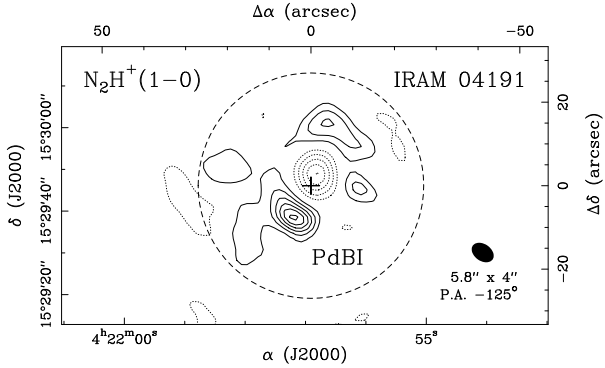}}}
 \vspace*{-1.0ex}
 \caption[]{N$_2$H$^+$(1-0) integrated intensity map of the IRAM~04191 inner 
   envelope obtained with PdBI (uncorrected for primary beam 
   attenuation). The emission is integrated over three velocity intervals 
   covering the whole seven-component multiplet. The contours vary from -0.5 
   to \hbox{-0.1} (dotted contours) and from 0.1 to 0.6 (solid  contours), by 
   step of 0.1 Jy~beam$^{-1}$~km~s$^{-1}$ (i.e. 3 times the rms
   noise level). The $54\arcsec$ (FWHM) primary beam is shown as a big circle
   and the synthesized clean beam (HPBW) is in the bottom right corner.
   The cross marks the central 1.3mm continuum position 
   of IRAM~04191 (cf. \citeauthor*{Belloche02}). 
   }
 \label{fig_n2h+map_pdbi}
\end{figure}

\begin{figure}[!t]
 \centerline{\resizebox{0.9\hsize}{!}{\includegraphics[angle=270]{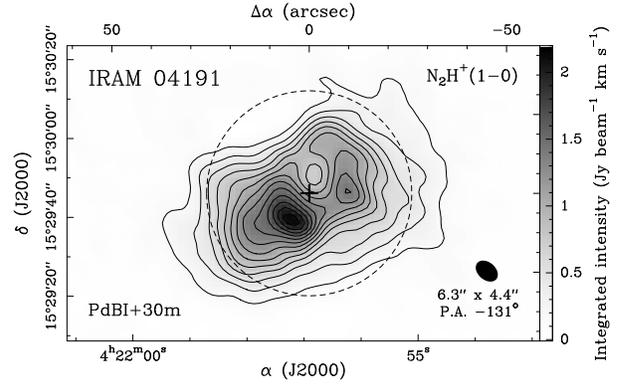}}}
 \vspace*{-1.0ex}
 \caption{N$_2$H$^+$(1-0) integrated intensity map of the IRAM~04191
   envelope combining PdBI and 30m data. The emission is integrated over the 
   same velocity intervals 
   as in Fig.~\ref{fig_n2h+map_pdbi}. The contours vary from 0.16 to 2.1 
   by step of 0.16 Jy~beam$^{-1}$~km~s$^{-1}$ (i.e. 6 times the rms
   noise). The map is {\it not} corrected for primary beam attenuation 
   (automatically applied to the single-dish data in the combination process); 
   the big circle represents the FWHM primary beam of the PdBI at 3mm.
   The synthesized clean beam (HPBW) is shown in the bottom right corner. 
   The cross marks the central position of IRAM~04191. 
   }
 \label{fig_n2h+map_pdbi30m}
\end{figure}

The N$_2$H$^+$(1-0) integrated intensity map obtained with PdBI shows two 
emission
peaks to the South-East and North-West of the central IRAM~04191 position 
(see Fig.~\ref{fig_n2h+map_pdbi}). 
An emission gap with negative contours down to the -15$\sigma$ level 
is also seen close to the central position, shifted by $3.5\arcsec$ (i.e., 
less than one beam width) to the North-West. 
These three features, which form a structure elongated roughly perpendicular 
to the outflow axis, have the same relative intensities in each $\Delta F_1$ 
group of hyperfine components of the N$_2$H$^+$(1-0) multiplet.
To recover the extended emission filtered out by the interferometer and obtain 
reliable N$_2$H$^+$ column densities, we 
combined our PdBI observations with short-spacing data obtained with the 30m 
telescope (\citeauthor*{Belloche02}). The resulting map of N$_2$H$^+$(1-0) 
integrated intensity reveals a ring-shaped core with two local peaks, 
surrounding an emission gap near the center (see 
Fig.~\ref{fig_n2h+map_pdbi30m}). This gap is a factor of 3.5 (2.3) weaker  
than the primary (secondary) emission peak to the South-East (West).
The detection of such a gap strongly suggests that the N$_2$H$^+$ abundance 
decreases toward the center. 

\subsection{Optical depths and excitation temperatures}
\label{sub_n2h+_tautex}

We used the hyperfine structure fitting method of the CLASS software 
to estimate the optical depths and excitation temperatures of N$_2$H$^+$(1-0) 
in the PdBI-30m combined map. The fits were limited to the spectra with a 
signal-to-noise ratio larger than 10. We fitted only the 4 
($\Delta F_1 = \pm1$) components to avoid convergence problems that occurred 
for a few positions in the western part of the envelope, and may be related to 
excitation anomalies \citep[e.g.][]{Turner01} and/or low signal-to-noise ratio 
for the weakest (110-011) component. The 1-$\sigma$ optical depth uncertainty 
given by the fits is typically 15-25 $\%$ for spectra with 
opacities larger than 5, and 50-150 $\%$ for optically-thinner spectra.
Except for a few positions at $\sim$ ($0\arcsec$, $15\arcsec$), 
the total N$_2$H$^+$(1-0) optical depths are lower than 15, and even lower 
than 10 toward the brightest emission peak and the emission gap, corresponding 
to optical depths lower than 1.7 and 1.1, respectively, for the isolated 
(101-012) component. 
The excitation temperature was found to vary from 4~K to 10~K across the 
PdBI-30m combined map and to display a maximum toward the brightest 
emission peak of the integrated intensity map. 

\subsection{Comparison with the dust emission maps}
\label{sub_n2h+_1.3mm_ratio}

\begin{figure}[!t]
 \centerline{\resizebox{0.9\hsize}{!}{\includegraphics[angle=270]{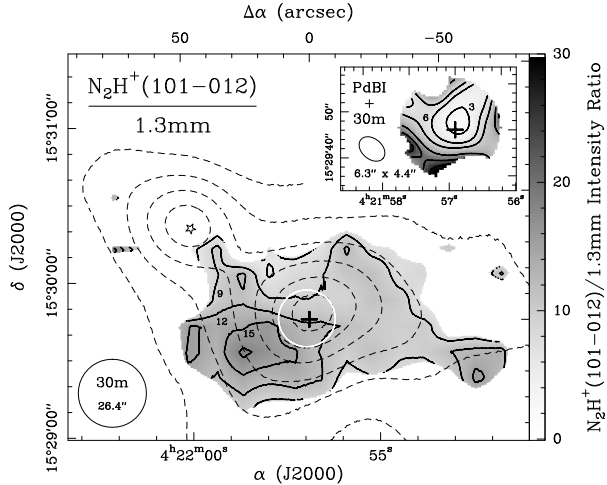}}}
 \vspace*{-1.0ex}
 \caption{Greyscale map of the ratio of N$_2$H$^+$(101-012) integrated 
   intensity to 1.3mm continuum intensity, both observed with the 30m 
   telescope. The (thick) contours go from 6 to 18 by 3 
   Jy\,beam$^{-1}$\,km\,s$^{-1}$/(Jy\,beam$^{-1}$). The 1.3mm continuum map 
   smoothed to the N$_2$H$^+$(1-0) resolution is overlaid as dashed contours.
   The insert in the top-right corner shows the PdBI-30m combined map 
   of the same ratio, limited to the 22$\arcsec$ PdBI primary beam at 227 GHz 
   (corresponding to the white circle in the larger map). The greyscale is 
   the same as for the larger map; the contours go from 3 to 9 by 3 and
   from 15 to 27 by 6 Jy\,beam$^{-1}$\,km\,s$^{-1}$/(Jy\,beam$^{-1}$).
   The beam sizes are shown in both maps. The cross marks the position of 
   IRAM~04191, the star indicates the  position of the Class I object  
   IRAS~04191.}
 \label{fig_n2h+_1.3mm_ratio}
\end{figure}

Figure~\ref{fig_n2h+_1.3mm_ratio} shows the ratio 
of N$_2$H$^+$(101-012) integrated intensity to 1.3mm continuum intensity 
across the IRAM~04191 envelope. The large greyscale map 
was obtained using the single-dish 1.3mm data 
of \citeauthor*{Motte01} smoothed to the $26.4\arcsec$ (FWHM) resolution
of the 30m telescope at the N$_2$H$^+$(1-0) frequency. 
The ratio was computed only at positions where both the 1.3mm and 
N$_2$H$^+$(101-012) intensities were stronger than 4 times the rms noise.
The higher-resolution, smaller greyscale map in the insert 
of Fig.~\ref{fig_n2h+_1.3mm_ratio} displays the same N$_2$H$^+$(101-012)
to dust continuum ratio derived from the PdBI-30m combined data  
for both the N$_2$H$^+$(101-012) and 1.3mm continuum emission. The PdBI 
continuum data were recalibrated to the frequency of the 30m continuum data 
prior to combination (see Sect.~\ref{sec_obs}), and the 1.3mm combined map was 
smoothed to the resolution of the N$_2$H$^+$(1-0) combined map. 
The ratio of the two sets of combined data was computed after correction 
for the respective PdBI primary beam attenuations at 3mm and 1.3mm. 

The N$_2$H$^+$(1-0) integrated intensity is expected to primarily track the 
variations of N$_2$H$^+$ column density 
(see Fig.~\ref{fig_n2h+_profile}a). Indeed, N$_2$H$^+$(1-0) is 
thermalized for $n_{\mbox{\scriptsize H$_2$}} > 10^5$ cm$^{-3}$, and the 
gas/dust temperatures 
\citep[which are well coupled above this density --][]{Doty97} are in the 
range 6.5--10~K 
(cf. Fig.~7d of \citeauthor*{Belloche02}), i.e., in a regime where the
LTE line intensities depend only weakly on temperature.
The 1.3mm continuum intensity, on the other hand, should scale as the H$_2$ 
column density times the dust temperature times the dust opacity 
\citep*[see, e.g., Eq.~1 of][]{Motte98}. The bolometric luminosity of 
IRAM~04191 is so low ($L_{bol} \sim 0.15\, L_\odot $ -- see 
\citeauthor*{Andre99}) that the dust temperature is expected 
to vary by less than $50\%$ in the range of radii 400--2800~AU 
(see \citeauthor*{Belloche02}).  
The variations of the intensity ratio in the central $r<2800$ AU 
($r_\theta< 20\arcsec$) region around IRAM~04191 
should thus reflect the variations of the N$_2$H$^+$ abundance, 
to within a factor of $\sim 2$ if we 
account for a possible slight saturation of N$_2$H$^+$(101-012)
($\tau_{\footnotesize 101-012} < 1.7$, see Sect.~\ref{sub_n2h+_tautex}) and 
possible changes in the dust opacity.

The intensity ratio has an average value of  
$\sim 11$~Jy\,beam$^{-1}$\,km\,s$^{-1}$/(Jy\,beam$^{-1}$) 
in the central ($r_\theta< 20\arcsec$) region of the single-dish map, 
and drops to a minimum of 2.7~Jy\,beam$^{-1}$\,km\,s$^{-1}$/(Jy\,beam$^{-1}$) 
near the center of the high-resolution map.
We conclude that the N$_2$H$^+$ abundance \textit{averaged} along the 
line of sight decreases by a factor of $\sim 4$ from 
$r_\theta \sim 20\arcsec$ to $r_\theta \sim 5\arcsec$.

\subsection{Disappearance of N\/$_2$H\/$^+$ from the gas phase}
\label{sub_n2h+_disappear}

\begin{figure}[!t]
 \centerline{\resizebox{\hsize}{!}{\includegraphics[angle=270]{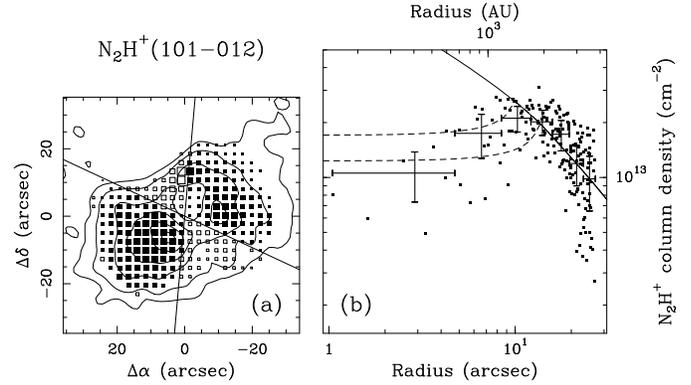}}}
 \vspace*{-1.0ex}
 \caption{
 \textbf{a)} Map of the N$_2$H$^+$ column density 
 $N_{\mbox{\scriptsize N$_2$H$^+$}}$ (squares) derived from the  
 N$_2$H$^+$(101-012) integrated intensity map overlaid as contours 
 (0.03, 0.06 to 0.3 by step of 0.06 Jy~beam$^{-1}$~km~s$^{-1}$). 
 The size of each square increases 
 linearly with $N_{\mbox{\scriptsize N$_2$H$^+$}}$, from 
 $2.6 \times 10^{12}$ to $3.3 \times 10^{13}$ cm$^{-2}$. 
 \textbf{b)} Circularly-averaged radial profile of 
 $N_{\mbox{\scriptsize N$_2$H$^+$}}$. The crosses with error bars represent 
 the average $N_{\mbox{\scriptsize N$_2$H$^+$}}$
 in concentric annuli limited to the positions shown with filled 
 squares in \textbf{a)}. The solid curve shows the H$_2$ column density 
 profile (\citeauthor*{Motte01}) scaled to the 7$^\mathrm{th}$ annulus 
 for comparison. The two dashed 
 curves illustrate the effect of a central hole of N$_2$H$^+$ material 
 of radius 1350~AU and 1850~AU, respectively.
 }
 \label{fig_n2h+_profile}
\end{figure}

The map of N$_2$H$^+$ column density shown in Fig.~\ref{fig_n2h+_profile}a 
was computed using a LTE partition function at the N$_2$H$^+$(1-0) 
excitation temperature derived from the hfs fits,
assumed to be uniform along each line of 
sight. We also applied an optical depth correction of $\tau/(1-e^{-\tau})$ 
using the hfs fit results. Figure~\ref{fig_n2h+_profile}b compares the 
circularly-averaged radial profile of N$_2$H$^+$ column density with the 
H$_2$ column density profile derived from the dust emission 
(\citeauthor*{Motte01}). 
The former clearly departs from the latter for $\theta \leq 13''$. A good 
model fit is obtained when one assumes a hole of N$_2$H$^+$ material 
in the central
part of the envelope ($r <$ 1350-1850 AU, see dashed curves). This
comparison strongly suggests that the N$_2$H$^+$ ion disappears from the
gas phase above a density of $n_{\mbox{\scriptsize H$_2$}} \sim 4-7 \times
10^5$ cm$^{-3}$ in the IRAM~04191 envelope, according to the 
density profile derived by \citeauthor*{Motte01} (see Fig.~7a of 
\citeauthor*{Belloche02}). We derive a mean N$_2$H$^+$ abundance 
of $\sim 7 \pm 4 \times 10^{-10}$ at $r = 3500$ AU.

\subsection{Fast, differential rotation}
\label{sub_rotation}

We measure a large velocity difference of $0.47 \pm 0.05$ km~s$^{-1}$ 
between the two N$_2$H$^+$(1-0) emission peaks of the PdBI map shown in 
Fig.\ref{fig_n2h+map_pdbi}, largely exceeding the FWHM 
linewidths measured toward the peaks themselves ($0.34 \pm 0.08$ km~s$^{-1}$).
A least-square linear fit to the LSR-velocity map of the PdBI-30m combined 
data yields a mean velocity gradient of 
$\sim 17$ km~s$^{-1}$pc$^{-1}$ with a position angle of P.A. $\sim 133^\circ$. 
This small-scale gradient has a 
direction consistent with the direction of the mean velocity gradient measured 
on larger scale with the 30m telescope
by \citeauthor*{Belloche02}, but is 5 times larger. The PdBI results 
thus confirm and strengthen the presence of fast, differential rotation 
identified by \citeauthor*{Belloche02}. 
The detailed form of the rotation velocity profile, which is confirmed to 
have a break at $r \simlt 3000$~AU (see Fig.~12 of \citeauthor*{Belloche02}),
will be discussed in a forthcoming paper
based on radiative transfer simulations taking the N$_2$H$^+$ depletion into 
account.

\section{Implications}

We showed in Sect.~\ref{sub_n2h+_disappear} that the molecular ion N$_2$H$^+$ 
disappears from the gas phase above a density of 
$n_{\mbox{\scriptsize H$_2$}} \sim 5 \times 10^5$ cm$^{-3}$ in the
envelope of the Class 0 protostar IRAM~04191. 
This conclusion is at variance with the results of recent 
chemical models \citep*[e.g.][]{Shematovich03,Aikawa03}. In particular, 
the models calculated by \citet*{Aikawa03} including gas-phase reactions, 
gas-dust interactions, and diffusive grain-surface reactions predict an 
{\it increase} in the N$_2$H$^+$ abundance during the prestellar collapse 
phase up to a density
of $n_{\mbox{\scriptsize H$_2$}} \sim 1 \times 10^7$ cm$^{-3}$ or more. 
IRAM~04191 is so young, probably less than $3 \times 10^4$~yr 
since the formation of the central object (\citeauthor*{Andre99}), 
that it is difficult to invoke a large decrease in the N$_2$H$^+$ abundance 
on such a short timescale at the beginning of the main accretion phase.

Since CO is one of the main destroyers of N$_2$H$^+$ in the gas phase 
\citep[e.g.][]{Aikawa01}, its potential desorption from grain mantles due 
to heating by the central protostar could be a mechanism responsible for
the disappearance of N$_2$H$^+$ from the gas phase. However, the 
luminosity of IRAM~04191 is so low (see 
Sect.~\ref{sub_n2h+_1.3mm_ratio}) that it is 
insufficient to raise the dust temperature above 11~K for $r > 400$ AU 
(J. Bouwman 2003, private communication). 
Comparing 
the C$^{18}$O(2-1) data of \citeauthor*{Belloche02}
with the 1.3mm continuum map of \citeauthor*{Motte01}, we
measure a reduction of the C$^{18}$O abundance (averaged along the line of 
sight) by a factor of $\sim 3.5$ 
toward the center of the envelope, assuming an optically thin 
C$^{18}$O line\footnote{Using our C$^{17}$(1-0), C$^{18}$(1-0) and 
C$^{18}$(2-1) data, we estimate the C$^{18}$(2-1) optical depth to be 
0.7-1 toward IRAM~04191.}. There is thus no evidence for any C$^{18}$O 
(and thus CO) desorption from the dust grains in the inner 
envelope at the 30m resolution (800 AU in radius). 
This indicates 
that potential CO desorption due to shocks produced by the outflow does not 
dominate in the central beam. Although the 
overall morphology seen in Figs.~\ref{fig_n2h+map_pdbi30m} and 
\ref{fig_n2h+_profile}a suggests that the N$_2$H$^+$ abundance may reach 
a minimum along the outflow path, the outflow is sufficiently well collimated 
(cf. Fig.~1 of \citeauthor*{Andre99}) that its effect should be negligible in 
the equatorial plane of the envelope.

We suggest two other possible explanations for the disappearance of 
N$_2$H$^+$ in the IRAM~04191 envelope:\\ 
1) N$_2$ may condense onto the grain surfaces. This could be either 
due to a higher N$_2$ binding energy on polar ice mantles than the value of 
750 K usually adopted \citep[see Model C of][]{Aikawa01,Bergin02}, or to an
enhanced grain chemistry transforming N$_2$ into less volatile species 
(Y. Aikawa 2003, private communication). 
Such a strong N$_2$ depletion was suggested by 
\citet{Caselli03} to account for the high H$_2$D$^+$ abundance measured in 
L1544.\\
2) Deuterium fractionation may enhance N$_2$D$^+$ over N$_2$H$^+$. A 
high-degree of deuterium fractionation has indeed been observed in dense
cores \citep[e.g.][]{Turner01,Bacmann03}, and \citet{Roberts03} predict 
that the N$_2$D$^+$/N$_2$H$^+$ abundance ratio should even exceed 1 
at densities of $\sim 10^6$ cm$^{-3}$ when they include HD$_2^+$ and D$_3^+$ 
in their chemical model. N$_2$D$^+$ may thus be a better tracer 
of dense gas than N$_2$H$^+$.

The disappearance of N$_2$H$^+$ from the gas phase suggests that all 
species bearing heavy elements also vanish at high density in 
prestellar cores. However, \citet*{Walmsley04} find that, under such 
conditions, the ionization degree of the gas
is not significantly changed owing to the enhanced abundances 
of H$^+$, H$_3^+$, and D$_3^+$. Moreover, they argue that the ambipolar 
diffusion timescale is dominated by the coupling of the neutrals to the 
charged grains at these densities.
The dynamics of protostellar collapse may thus be essentially 
unaffected by the strong molecular depletion in prestellar cores.

\begin{acknowledgements}
 We would like to thank Y. Aikawa, M. Gerin, G. Pineau des For\^ets and 
 M. Walmsley for enlightening discussions about the chemistry. We are grateful 
 to F. Gueth, J. Pety and R. Neri (IRAM) for their help with the Plateau de 
 Bure observations.  We would also like to thank F. Motte for 
 her advice about the continuum data.
\end{acknowledgements}

\end{document}